# Structural Characterization of Magnetoferritin

[a]Lucia Melníková, [a]Zuzana Mitróová, [a]Milan Timko, [a]Jozef Kováč, [b]Mikhail V. Avdeev, [b,c]Viktor I. Petrenko, [d]Vasyl M. Garamus, [e]László Almásy, [a]Peter Kopčanský

[a]*Institute of Experimental Physics, SAS, Watsonova 47, 040 01 Kosice, Slovakia*
[b]*Frank Laboratory of Neutron Physics, Joint Institute for Nuclear Research, Dubna, Russia*
[c]*Taras Shevchenko National University of Kyiv, Kyiv, Ukraine*
[d]*Helmholtz-Zentrum Geesthacht, Zentrum für Material- und Küstenforschung, Geesthacht, Germany*
[e]*Wigner Research Centre for Physics, Institute for Solid State Physics and Optics. H-1525, Budapest, P. O. Box 49, Hungary*

The physicochemical characterization of the magnetoferritin biomacromolecule in terms of morphology, structural and magnetic properties shows that iron oxides can be efficiently loaded into apoferritin molecules, preserving their native, bio-compatible structure and affecting the morphology of the protein shell.

Natural ferritin is the iron-storage protein of animals, plants, and bacteria. It is a spherical biomacromolecule of external diameter about 12 nm composed of 24 protein subunits arranged as a hollow sphere of approximately 8 nm in diameter. Inside the sphere, iron is stored in the ferric oxidation state as complex molecule with a crystallographic structure similar to the mineral ferrihydrite [1]. By a suitable chemical process, magnetic iron oxide nanoparticles ($Fe_3O_4$, $\gamma-Fe_2O_3$) can be synthesized in the empty protein shell of ferritin, i.e. apoferritin, forming a biocompatible ferrofluid, called magnetoferritin [2,3]. The problem of toxicity and side effects of magnetic nanoparticles in organs and tissues is minimized due to the protein nature of this material, which is important for many possible applications in cell labeling, biological separation and clinical practice. Their magnetic properties, based on their inducible magnetization, allow them to be heated by externally applied AC magnetic field. It makes them attractive for many applications, ranging from various magnetic separation techniques and contrast enhancing agents for MRI to magnetic hyperthermia [4, 5]. Magnetoferritin is a promising compound which can be used as a drug carrier; the protein shell is able bind to tumor cells via transferin receptor 1 (TfR1) [6] and the drug can be bound to the protein subunit. In addition to biocompatibility, another advantage for biotechnological applications of magnetoferritin is a relatively short time of controlled synthesis [7, 8].
The main interest of the present study is associated with understanding of certain diseases development mechanism that is in a close relationship with the iron metabolism and iron storage protein, ferritin [9]. In healthy organisms, ferritin is able to store up to 4500 Fe atoms in a ferrihydrite-like mineral core [10]. Many researchers confirmed the presence of magnetite nanoparticles inside pathological tissues [11, 12] which is related to the $Fe^{2+}$ ions accumulation and defects in the normal storage function of ferritin [13]. This indicates the transformation of ferrihydrite to magnetite and formation of biogenic magnetoferritin. The precise mode of such transformation regulated by the biochemistry of organisms (presence of specific enzymes, bio-complexes, etc.) has not been determined yet. Therefore, in this research, magnetoferritin prepared by *in vitro* chemical synthesis was used as a model system of pathological ferritin. Structural studies of ferritin and magnetoferritin would be useful to elucidate the structural changes of ferritin shell disruption or aggregation which is observed in development of many cancer or neurodegenerative diseases [14, 15].

In this study, magnetoferritin prepared by controlled chemical synthesis, in accordance with a procedure described previously [16] was the test material characterized by dynamic light scattering (DLS), transmission electron microscopy (TEM), small-angle neutron and X-ray scattering (SANS, SAXS) and SQUID magnetometry.

The zeta potentials of the test apoferritin and magnetoferritin at comparable concentrations of 2.11 mg/ml and 2.36 mg/ml were -25.5 mV, -21.9 mV respectively. The results confirm the negative charge of the molecule and its good stability. The hydrodynamic diameter of magnetoferritin molecules was measured and compared with that of apoferritin by dynamic light scattering technique with protein concentration of 0.3 g.L$^{-1}$ in both solutions (Fig. 1.).

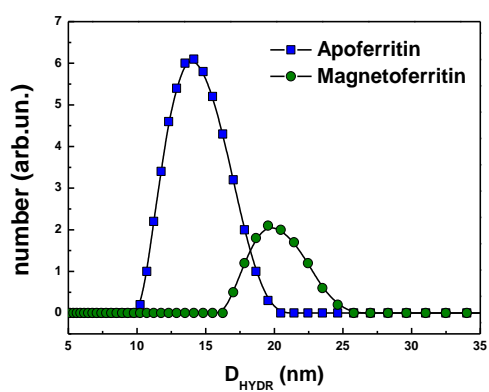

**Fig. 1.** The size distributions of apoferritin and magnetoferritin as revealed by DLS

Generally, the hydrodynamic diameter is larger than the theoretical size because it indicates the effective size of the hydrated/solvated molecule. In comparison with apoferritin hollow sphere (<$D_{HYDR}$> = 14.14 nm), the hydrodynamic diameter of magnetoferritin increases (<$D_{HYDR}$> = 19.54 nm). Such increase can be related to a deformation of the particles upon loading with iron oxide and to the presence of some fraction of aggregated particles.

Transmission electron microscopy showed presence of well-defined rounded nanocrystallites (Fig. 2.) with average diameter of 5 nm.

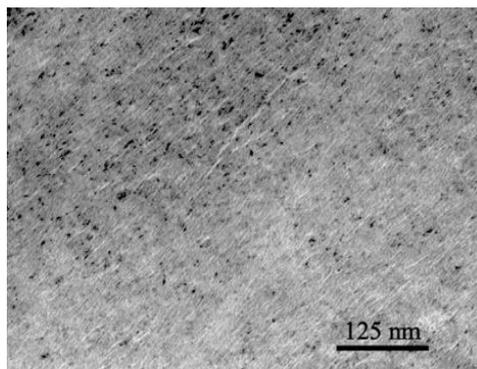

**Fig. 2.** TEM image of magnetoferritin

The electron diffraction of magnetoferritin samples confirmed the face-centered cubic crystalline structure of the ferrous phase but it is not possible distinguish between magnetite ($Fe_3O_4$) or maghemite ($\gamma-Fe_2O_3$). More information can be obtained by magneto-optical birefringence or Faraday rotation studies as was shown in our recent works [17,18].

Small-angle neutron scattering measurements were performed at the Yellow Submarine instrument operating at the Budapest Neutron Centre [19]. Samples were prepared by redispersing apoferritin and magnetoferritin in $D_2O$ from dry powder to form a 2 wt. percent solution.

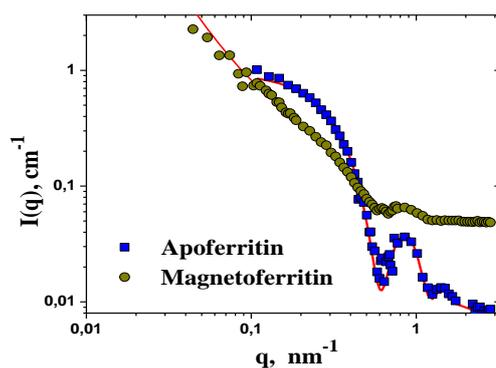

**Fig. 3.** SANS data of magnetoferritin and apoferritin dispersions in $D_2O$. The solid lines are model fits of a spherical shell to the apoferritin data and the same model including aggregated particles for the magnetoferritin data.

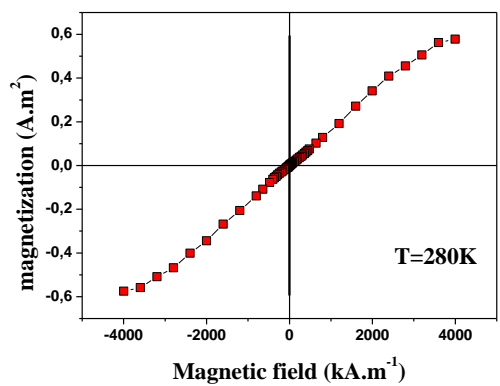

**Fig. 4.** Field dependence of magnetoferritin magnetization measured at room temperature.

Fig. 3. compares the scattering data from pure (unloaded) apoferritin with the scattering of magnetoferritin. For both solutions, the minima and maxima characteristic to the spherical shell form factor of apoferritin are seen. For magnetoferritin, the oscillations are less pronounced, indicating that the spherical form of the protein is only partly preserved. The relative weakening of the characteristic shell structure is attributed to a deformation of the protein shell upon loading, leading to decrease in the spherical symmetry of the molecules. The increase in the scattering intensity at small values of q shows that a fraction of the molecules aggregate to loose objects of sizes larger than 200 nm. The experimental data were modeled using the hollow spherical shell model of apoferritin and aggregated spherical shell particles for the case of magnetoferritin. The solution contains both aggregated and non-aggregated particles, and the used modeling could not distinguish these populations; therefore the fraction of aggregated particles could not be extracted from the data. Small-angle X-ray scattering data, taken using a laboratory setup, confirmed that the aggregates contain iron oxide.

Magnetic properties of magnetoferritin were investigated using SQUID magnetometer in magnetic fields up to 4 000 $kA.m^{-1}$. The samples show superparamagnetic behavior without hysteresis at room temperature (Fig. 4). Using the particle size as obtained by TEM, and assuming magnetite, the saturation magnetization of 8 $A.m^2.kg^{-1}$ was calculated. The observed magnetization is by an order of magnitude lower than this value, indicating that the magnetic core of magnetoferritin presumably consists of mixed hematite and magnetite. The magnetization curves measured at 2K below blocking temperature ($T_b$ = 26 K) showed the hysteresis with coercive field of 20.0 $kA.m^{-1}$. The magnetization measured at 5 K undergoes a slow approach to saturation at field which we can achieve.

In conclusion, the synthesized materials demonstrate a superparamagnetic behavior, the structure determined by TEM and scattering shows that magnetic nanoparticles are confined in the spherical protein shell with particle diameters about 5 nm, thus, not filling the entire available space. The protein structure slightly changes upon loading, and this change can be attributed to the effect of iron oxides binding and ordering inside the protein cavity of magnetoferritin. Further experiments, for example, contrast variation SANS methods would give more detailed information concerning the protein and the magnetic structure of magnetoferritin with different loading factors, to reveal how the iron oxides affect protein conformation. Clarification of these effects could have a major impact in biomedicine for understanding the role of magnetite in connection with aggregation process in the development of neurodegenerative diseases.

*Acknowledgement(s). This work was supported by the projects Nos. 26220120021, 26220220005 and 26110230061 in the frame of Structural Funds of European Union, Centre of Excellence of SAS Nanofluid and VEGA 0041, 0045 as well as the Slovak Research and Development Agency under the Contract No APVV 0171-10. The neutron scattering experiments have been supported by the European Commission under the 7th Framework Program through the Key Action: Strengthening the European Research Area, Research Infrastructures (Grant Agreement no. 283883 NMI3). L.A. acknowledges the support of the Hungarian Scholarship Board for a short research stay at the IEP SAS.*